# FeCl$_3$ based Few-Layer Graphene Intercalation Compounds: Single Linear Dispersion Electronic Band Structure and Strong Charge Transfer Doping

By *Da Zhan, Li Sun, Zhenhua Ni\*, Lei Liu, Xiaofeng Fan, Yingying Wang, Ting Yu, Yeng Ming Lam, Wei Huang,  Zexiang Shen\**

[*]   Prof. Z. X. Shen, D. Zhan, L. Sun, Dr. L. Liu, Dr. X. F. Fan, Dr. Y. Y. Wang, Prof. T. Yu
Division of Physics and Applied Physics, Nanyang Technological University
Singapore 637371 (Singapore)
E-mail: zexiang@ntu.edu.sg
      Prof. Z. H. Ni
Department of Physics, Southeast University,
Nanjing 211189, (P. R. China)
E-mail: zhni@seu.edu.cn
      Prof. Y. M. Lam
School of Materials Science and Engineering, Nanyang Technological University
Singapore 639798 (Singapore)
       Prof. W. Huang
Institute of Advanced Materials, Nanjing University of Posts and Telecommunications, 9 Wenyuan Road
Nanjing 210046, (P. R. China)

Graphene has attracted great attentions since its first discovery in 2004. Various approaches have been proposed to control its physical and electronic properties. Here, we report that graphene based intercalation compounds is an efficient method to modify the electronic properties of few layer graphene (FLG). FeCl$_3$ intercalated FLG were successfully prepared by two-zone vapor transport method. This is the first report on full intercalation for graphene samples. The features of the Raman G peak of such few-layer graphene intercalation compounds (FLGIC) are in good agreement with their full intercalation structures. The FLGIC presents single Lorentzian 2D peak, similar to that of single layer graphene, indicating the loss of electronic coupling between adjacent graphene layers. First principle calculations further reveal that the band structure of FLGIC is similar to single layer graphene but with strong doping effect due to the charge transfer from graphene to FeCl$_3$. The successful fabrication of FLGIC opens a new way to modify properties of FLG for fundamental studies and future applications.

# 1. Introduction

Graphene, a single layer of carbon atoms with hexagonal arrangement, has attracted enormous interest due to its excellent electric field effect transport properties[1] and massless Dirac Fermions like charge carriers.[2,3] Huge progress on graphene research has been achieved in recent years including: Various methods for graphene fabrication;[1,4-7] Discovery of its unique electronic, thermal and mechanical properties;[1-3, 8-11] Successful fabrication of prototype graphene-based devices.[12-13] However, much of the unique properties of graphene are accorded to that of single layer graphene (SLG). It would be very desirable to modify few-layer graphene (FLG) samples so that they have similar properties as that of SLG.

Graphite intercalation compounds (GICs) are complex materials that formed by insertion of atomic or molecular layers of different chemical species between graphite interlayer space.[14] The interlayer distance of graphite is dramatically increased due to presence of intercalants, which would strongly affect the electronic coupling between graphene layers, hence changes its properties. Moreover, due to the wide variation of intercalants, different physical properties can be achieved for GIC, such as electrical, thermal and magnetic properties.[14-16] Thus, graphene based intercalation compounds would be an efficient method to modify the properties of FLG. Until now, there are no experimental reports on few-layer graphene intercalation compounds (FLGIC) except the most recent report on $Br_2$ and $I_2$ intercalated FLG by Jung et al., where FLG is not fully intercalated according to their Raman spectra and the structural model.[17] In this work, fully intercalated $FeCl_3$-FLGIC have been successfully prepared and systematically investigated by Raman spectroscopy. Raman spectra of such FLGIC clearly reveal the single-layer graphene like electronic structure and strong charge transfer induced doping effect. First principle calculations are also carried out to confirm the experimental results.

## 2. Result and Discussion

### 2.1. Full intercalation structure of FLGIC revealed by Raman G Peak Features

Stage number is a key factor for normal bulk based GIC, where stage is defined as the number of graphene layers between the adjacent intercalated layers.[18] Raman spectroscopy is an effective tool to confirm the intercalation stage of GIC by identifying the component and structure of G peak, an $E_{2g}^{(2)}$ in plane vibrational mode of graphite. The frequency of G peak in GIC is known to be affected by three factors: charge transfer, intercalate-coupling effect and change of lattice constant.[19,20] The degree of charge transfer usually dominate the frequency evolution.[19,21] For the case of $FeCl_3$-GIC, the position and shape of the graphene G peak can differ under different intercalation conditions. For example, the singlet G peak position can blue shift to ~1612 cm$^{-1}$ for stage 2 GIC, representing graphene layer flanked on one side by $FeCl_3$, or blue shift to ~1626 cm$^{-1}$ for stage 1 GIC, representing graphene layer flanked on both sides by $FeCl_3$.[22,23] Furthermore, singlet G peak splits into doublet structure for GIC with higher stage (stage>2) or mixed stages.[22,23] The position and intensity ratio of each component of doublet peaks can be used to confirm the exact situation.

$FeCl_3$-FLGIC was fabricated by traditional two-zone method (see Experimental Section). Fig. 1 presents the Raman spectra for graphenes (1 layer (1L) to 4 layers (4L)) after intercalation. For SLG, it is not possible to be intercalated. Instead, we find doping induced stiffening and sharpening of the G peak (blue shifts from 1581 to 1604 cm$^{-1}$), which is quite normal for SLG after vacuum annealing and exposure to air.[24,25]

For 2L graphene intercalated by $FeCl_3$, the G peak position blue shifts to ~1612 cm$^{-1}$ ($G_1$) from ~1580 cm$^{-1}$ of pristine 2L graphene, which is similar to the previous reported G peak position of stage 2 bulk GIC.[22-23] Such a large blue shift (~32 cm$^{-1}$) in G peak cannot be due to substrate and interface charge density doping,[26-27] which normally only introduce a shift of less than 10 cm$^{-1}$. The similar Raman spectra of 2L-FLGIC and stage 2 GIC can be explained by their similarity in structures. For 2L-FLGIC, both graphene layers are flanked on one side by FeCl$_3$ layer (Fig. 2a), same as the structure of stage 2 GIC. Thus, it is reasonable that the G peak of 2L-FLGIC still presents one singlet peak and its position is similar to that of stage 2 bulk GIC.

For 3L and 4L graphene fully intercalated by FeCl$_3$, both of them present doublet G band, with one peak located at ~1612 cm$^{-1}$(G$_1$) and the other peak at ~1623 cm$^{-1}$(G$_2$). Fig. 2b and 2c show the schematic structures of 3L and 4L FLGIC, the top and bottom graphene layers (carbon atoms in yellow) are flanked on one side by FeCl$_3$ layer, and they contribute to the G$_1$ peak similar as in 2L-FLGIC, and hence the peak intensity of G$_1$ is almost same for 2L to 4L-FLGIC (Fig. 1). The middle graphene layer(s) (carbon atoms in orange) is(are) flanked on both sides by FeCl$_3$ layers, which is similar to the case of stage 1 FeCl$_3$ based bulk GIC,[22,23] and they contribute to the more blue shifted G$_2$ peak at ~1623 cm$^{-1}$. The integrated intensity ratio of I$_{G1}$/I$_{G2}$ is in the range of 1.2-1.8 and 0.4-0.8 for 3L-FLGIC and 4L-FLGIC, respectively, which are close to ratio of 2 and 1 according to the schematic crystal structure illustrated in Fig. 2b and 2c.

## 2.2. Strong Charge Transfer Chemical Doping

The stiffening of the E$_{2g}^{(2)}$ phonons (blue shift of G$_1$ and G$_2$ peaks) of graphene layer for FeCl$_3$-FLGIC is mainly because of charge transfer induced doping effect,[17] where the graphene layer can be considered as hole doped as FeCl$_3$ is acceptor type intercalant.[20] The strong charge transfer effect induces downshift of Fermi surface of graphene and causes stiffening of G peak due to the nonadiabatic removal of the Kohn anomaly at Γ point.[28-30] The larger blue shift of the G$_2$ peak compare with the G$_1$ peak is because the inner graphene layers are affected on both sides by the FeCl$_3$ layers (G$_2$) while the outer layers are only one side ($G_1$). In addition to stiffening effect of $E_{2g}^{(2)}$ phonons, the much smaller linewidths of $G_1$ (~7 cm$^{-1}$) and $G_2$ (~6 cm$^{-1}$) peaks compared to that of pristine graphene (~15 cm$^{-1}$) are also observed, which is another indication of strong doping effect on graphene.[28]

The G peak position of FLGIC samples versus the number of graphene layers is shown in Fig. 3. For comparison, G peak positions of pristine graphene samples as well as doped graphene samples are also included. As can be seen in Fig. 3, the G peak position of pristine graphene is almost independent of number of layers. For doped 1L graphene, obvious blue shifts from ~1580 cm$^{-1}$ to ~1605 cm$^{-1}$ was observed. The amount of blue shift is decreased with the increase of graphene layer number from 2L to 4L. On the other hand, for FLGIC, the phenomena are totally different. For 2L-FLGIC, the G peak blue shifts to 1612 cm$^{-1}$, and it splits into two peaks ($G_1$ and $G_2$) when the number of graphene layers increases to 3 and more. The $G_1$ peak intensity is almost constant in our experiments, while the $G_2$ peak intensity becomes stronger with the increase of the graphene layer number. The intensity of the peaks is represented by the size of the solid circles in Fig. 3. The trend of the $G_1$ and $G_2$ intensity evolution from 2L- to 4L-FLGIC infers that with the increase of the number of graphene layers, the $G_2$ peak will gradually dominate the doublet G band and finally it should present like a single peak as stage 1 bulk-based GIC case (~1626 cm$^{-1}$).[22,23]

**2.3. Electronic Band Structure of FLGIC Probed by Raman 2D Peak**

The 2D peak can be used as a fingerprint to identify layer numbers of pristine graphene, as it is originated from double resonance process and strongly dependent on the electronic band structure around K point at the Brillouin zone.[31] The 2D peak of fully intercalated FLGIC samples shows essential difference from that of pristine FLG in the following three aspects. (1) Only one single Lorentzian peak can be well fitted for all FLGIC samples (2L- to 4L- FLGIC) as shown in Fig. 1; this is similar to the 2D peak of SLG[32] as well as mis-oriented graphene or folded graphene.[33] In contrast, the 2D peak of pristine 2L graphene can be fitted by four

Lorentzian peaks and those of other FLG are fitted by many Lorentzian peaks.[31,32] (2) The linewidth of the 2D peak of FLGIC is much sharper than that of pristine FLG, as shown in Fig. 4. For pristine and doped graphene, the linewidth dramatically broadens from 1L to few layers. However, for FLGIC, the 2D linewidths are much sharper than FLG and they only show slight broadening with the increase of number of layers. (3) The 2D peak integrated intensity of pristine FLG does not change with the increased number of graphene layers.[32] However, for FLGIC, 2D peak intensity increases with number of layers (In Fig. 1, all the spectra were measured under same condition).

The above difference of 2D peak properties between FLGIC and FLG are mainly because of two reasons: (1) The distance between adjacent graphene layers is enlarged from original ~3.4 Å to ~9.4 Å in the FLGIC samples.[22,23,34] Thus, adjacent graphene layers can only interact with each other through the intercalant layer, resulting in much weaker coupling, which in turn make the properties of FLGIC very similar to that of SLG, as indicated by sharp SLG-like 2D peaks. (2) Graphene and $FeCl_3$ are incommensurate in structure because the lattice constant of $FeCl_3$ and graphene are 6.06 Å and 2.46 Å,[14,22,34] respectively. Therefore, the coupling between graphene and $FeCl_3$ is very weak. As a result, the fully intercalated $FeCl_3$-FLGIC can be viewed as quasi-individual graphene layers superimposed together with very weak coupling effect through the intercalant layers. The electronic properties of FLGIC behave like SLG with single dispersion near Dirac point, resulting in single Lorentzian 2D peak with peak intensity increases with number of layers.

## 2.4. Electronic Band Structure by First Principle Calculations

The calculated band structure of $FeCl_3$-based stage 1 GIC (blue solid curve) and SLG (red dashed curve) are shown in Fig. 5. The horizontal bands of GIC around 0 and 1.5 eV originate mostly from the *d* orbitals of iron. Except for those bands, the band structure of GIC is very similar to that of SLG, with single dispersion near Dirac point. This agrees well with the Raman spectra of FLGIC which show single and sharp 2D peak. Considering the on-site Coulomb interaction of $Fe^{3+}$ ions, the spin-dependent DFT calculation (LSDA+U) confirms the role of $FeCl_3$ insertion in FLG. While bringing little disturbance to the band of SLG, the main effect of the inserted $FeCl_3$ layers on SLG is shifting the Fermi level and transferring charges. The Fermi energy of FLGIC shifts to ~1.0 eV below the Dirac point, which indicates FLGIC is heavily hole doped. This hole doping effect induced by charge transfer from graphene to $FeCl_3$ explains the significant blueshift of G peak of FLGIC. The $G_2$ peak of FLGIC is at ~1623 $cm^{-1}$, which can be converted to the shift of Fermi level of 0.8-0.9 eV from the extrapolated gate controlled doping result.[28] This again matches the calculated value of 1.0 eV quite well.

**2.5. Homogenous Intercalation of $FeCl_3$ on FLGIC**

Fig. 6a shows the optical image of graphene sample before intercalation process. The 1L and 2L graphene were identified by Raman and contrast spectra before intercalation. Fig. 6b and 6c, respectively, show Raman images of the G peak and 2D peak intensity after intercalation. The intensities of G and 2D peaks of 2L-FLGIC are much stronger than those of 1L graphene. However, the 2D peak linewidth does not show any noticeable difference between 1L graphene and 2L-FLGIC as shown in Fig. 6d. All the observed phenomena are consistent with the results in Fig. 1 and 4. The uniform 2D linewidth across the whole 2L-FLGIC region indicates the homogenous intercalation of $FeCl_3$. $FeCl_3$-FLGIC samples are very stable in air ambient and the G and 2D Raman peaks remain unchanged for 3 months until now. This result is similar to the bulk based $FeCl_3$-GIC.[35] In addition, no obvious D peak was observed for FLGIC indicating the good crystalline quality of graphene samples after intercalation.

Compared with the $Br_2$ or $I_2$ based FLGIC reported recently,[17] FLGs are easier to be fully intercalated by $FeCl_3$. This is not surprising as $Br_2$ usually creates stage 2 bulk GIC.[36] The observed complete quenching effect of 2D band in Br$_2$-FLGIC might be due to the commensurate structure between Br$_2$ and graphene layers, resulting in stronger interaction and modification of the electronic structure of graphene.[17,37] On the other hand, FeCl$_3$ is incommensurate with graphene and hence the interaction between FeCl$_3$ and graphene is minimum. The main effect of FeCl$_3$ intercalation is to: 1. Introduce strong charge transfer doping. 2. Increase the effective distance between the graphene layers and thus the electronic structure of FeCl$_3$-based FLGIC becomes similar to that of SLG. Such effects could be further demonstrated by electrical transport measurements, where a significant shift of neutral point (Dirac point) as well as characteristics of SLG (i.e. Berry's phase $\pi$)[38] should be present on FLGIC. Therefore, the electrical transport properties of FLGIC are presumably very interesting and deserved for further studies. The different effects of FLGIC using different intercalants suggest FLGIC offers a promising method to achieve desirable properties for FLG samples for both fundamental research and future applications.

## 3. Conclusions

In summary, air ambient stable FeCl$_3$-FLGIC samples with homogenous concentration of intercalant have been successfully prepared. Our results show that FLG samples are much easier to be fully intercalated compared with bulk graphite, which normally takes ~6 days in high concentration Cl$_2$ atmosphere.[39] Raman spectroscopy and imaging confirm that FeCl$_3$ is fully intercalated into FLG, while simultaneously introduces strong charge transfer chemical doping. FLGIC show single and sharp 2D peak, similar to that of single layer graphene, indicating the loss of electronic coupling between adjacent graphene layers. The observed phenomena agree very well with first principle calculations. FLGIC are quite promising materials not only because the modification of the graphene electronic structure, but also the possible modification of electrical, thermal, and magnetic properties with choice of different intercalants, particularly the Ca-FLGIC, which is expected to show superconductivity.[40]

## 4. Experimental

*Fabrication of pristine FLG, doped FLG and FLGIC*: Pristine graphene sheets were deposited by mechanical cleavage on silicon wafer covered by 300 nm thick $SiO_2$. The number of graphene layers (1 layer (1L) to 4 layers (4L)) was confirmed by Raman spectroscopy and their contrast spectra.[41,42] The two-zone vapor transport method was used for fabricating $FeCl_3$-FLGIC. The reaction took place inside a vessel constructed from glass tube. The graphene samples and anhydrous $FeCl_3$ powder (~0.03 gram) were separated by ~6 cm. The tube was pumped to $10^{-2}$ torr and sealed. In our experiment, two-zone method was processed using single furnace instead of traditional two-furnace technique. The temperature distribution in furnace is measured by a thermocouple before experiment. The reaction vessel was placed inside the proper position of furnace to achieve our desired two-zone temperatures (360℃ for graphene samples and 310℃ for anhydrous $FeCl_3$ powder). The heating rate was set as 10℃/min, and the vessel was kept for 10 hours at the setting temperatures. Finally the furnace was cooled with a rate of 10℃/min to room temperature. The doped graphene is prepared by vacuum annealing of graphene at 900 ℃ for 10 min and then exposed to air ambient at room temperature to obtain molecular ($H_2O/O_2$) doping.[24,25]

*Raman Spectroscopy Measurement*: Raman spectra were recorded by WITEC CRM200 system with spectral resolution of 1 $cm^{-1}$. The excitation laser is 532 nm (2.33 eV) and the laser power at sample is kept below 0.5 mW to avoid laser heating effect. A 100× objective lens with a numerical aperture of 0.95 was used. For Raman mapping measurement, a piezostage was used to move the sample with step size of 250 nm and Raman spectrum is recorded at every point in the scanned area.

*Electronic Band Structure Calculation*: The structure of $FeCl_3$-based stage 1 GIC modeled by M. S. Dresselhaus et al.[14,22,34] is simulated with the density functional theoretical (DFT) calculation. A supercell with lattice constants a=12.12Å and c= 9.370Å is constructed, where the layered $FeCl_3$ with 2×2 periods is taken as commensurate with the graphene of 5×5 unit cells as shown in Fig. 3. The DFT calculations based on the generalized gradient approximation (PBE-GGA)[43] are performed using the plane-wave basis VASP code.[44] The projector augmented wave (PAW) method is employed to describe the electron-ion interactions. A kinetic energy cutoff of 400 eV and k-points sampling with 0.05Å$^{-1}$ separation in the Brillouin zone are used. The structure is optimized by a conjugate gradient algorithm with a force convergence criterion of 0.01 eV/Å.

# References

[1] K. S. Novoselov, A. K. Geim, S. V. Morozov, D. Jiang, Y. Zhang, S. V. Dubonos, I. V. Grigorieva, A. A. Firsov, *Science* **2004**, *306*, 666.
[2] K. S. Novoselov, A. K. Geim, S. V. Morozov, D. Jiang, M. I. Katsnelson, I. V. Grigorieva, S. V. Dubonos, A. A. Firsov, *Nature* **2005**, *438*, 197.
[3] Y. B. Zhang, Y. W. Tan, H. L. Stormer, P. Kim, *Nature* **2005**, *438*, 201.
[4] C. Berger, Z. M. Song, T. B. Li, X. B. Li, A. Y. Ogbazghi, R. Feng, Z. T. Dai, A. N. Marchenkov, E. H. Conrad, P. N. First, W. A. de Heer, *J. Phys. Chem. B* **2004**, *108*, 19912.
[5] S. Stankovich, D. A. Dikin, G. H. B. Dommett, K. M. Kohlhaas, E. J. Zimney, E. A. Stach, R. D. Piner, S. T. Nguyen, R. S. Ruoff, *Nature* **2006**, *442*, 282.
[6] A. Reina, X. T. Jia, J. Ho, D. Nezich, H. B. Son, V. Bulovic, M. S. Dresselhaus, J. Kong, *Nano Lett.* **2009**, *9*, 30.
[7] D. V. Kosynkin, A. L. Higginbotham, A. Sinitskii, J. R. Lomeda, A. Dimiev, B. K. Price, J. M. Tour, *Nature* **2009**, *458*, 872.
[8] S. Ghosh, I. Calizo, D. Teweldebrhan, E. P. Pokatilov, D. L. Nika, A. A. Balandin, W. Bao, F. Miao, C. N. Lau, *Appl. Phys. Lett.* **2008**, *92*, 151911.
[9] S. V. Morozov, K. S. Novoselov, M. I. Katsnelson, F. Schedin, D. C. Elias, J. A. Jaszczak, A. K. Geim, *Phys. Rev. Lett.* **2008**, *100*, 016602.
[10] J. H. Chen, C. Jang, S. D. Xiao, M. Ishigami, M. S. Fuhrer, *Nat. Nanotechnol.* **2008**, *3*, 206.
[11] A. A. Balandin, S. Ghosh, W. Z. Bao, I. Calizo, D. Teweldebrhan, F. Miao, C. N. Lau, *Nano Lett.* **2008**, *8*, 902.
[12] M. D. Stoller, S. J. Park, Y. W. Zhu, J. H. An, R. S. Ruoff, *Nano Lett.* **2008**, *8*, 3498.
[13] X. Wang, L. J. Zhi, K. Mullen, *Nano Lett.* **2008**, *8*, 323.
[14] M. S. Dresselhaus, G. Dresselhaus, *Adv. Phys.* **2002**, *51*, 1.
[15] T. E. Weller, M. Ellerby, S. S. Saxena, R. P. Smith, N. T. Skipper, *Nat. Phys.* **2005**, *1*, 39.
[16] N. Emery, C. Herold, M. d'Astuto, V. Garcia, C. Bellin, J. F. Mareche, P. Lagrange, G. Loupias, *Phys. Rev. Lett.* **2005**, *95*, 087003.
[17] N. Jung, N. Kim, S. Jockusch, N. J. Turro, P. Kim, L. Brus, *Nano Lett.* **2009**, *9*, 4133.
[18] P. C. Eklund, D. S. Smith, V. R. K. Murthy, S. Y. Leung, *Synth. Met.* **1980**, *2*, 99.
[19] C. T. Chan, K. M. Ho, W. A. Kamitakahara, *Phys. Rev. B* **1987**, *36*, 3499.
[20] T. Abe, M. Inaba, Z. Ogumi, Y. Yokota, Y. Mizutani, *Phys. Rev. B* **2000**, *61*, 11344.
[21] C. T. Chan, W. A. Kamitakahara, K. M. Ho, P. C. Eklund, *Phys. Rev. Lett.* **1987**, *58*, 1528.
[22] N. Caswell, S. A. Solin, *Solid State Commun.* **1978**, *27*, 961.
[23] C. Underhill, S. Y. Leung, G. Dresselhaus, M. S. Dresselhaus, *Solid State Commun.* **1979**, *29*, 769.
[24] L. Liu, S. M. Ryu, M. R. Tomasik, E. Stolyarova, N. Jung, M. S. Hybertsen, M. L. Steigerwald, L. E. Brus, G. W. Flynn, *Nano Lett.* **2008**, *8*, 1965.
[25] Z. H. Ni, H. M. Wang, Z. Q. Luo, Y. Y. Wang, T. Yu, Y. H. Wu, Z. X. Shen, *J. Raman Spectrosc.* **2010**, *41*, 479.
[26] I. Calizo, W. Z. Bao, F. Miao, C. N. Lau, A. A. Balandin, *Appl. Phys. Lett.* **2007**, *91*, 201904.
[27] Y. Y. Wang, Z. H. Ni, T. Yu, Z. X. Shen, H. M. Wang, Y. H. Wu, W. Chen, A. T. S. Wee, *J. Phys. Chem. C* **2008**, *112*, 10637.
[28] A. Das, S. Pisana, B. Chakraborty, S. Piscanec, S. K. Saha, U. V. Waghmare, K. S. Novoselov, H. R. Krishnamurthy, A. K. Geim, A. C. Ferrari, A. K. Sood, *Nat. Nanotechnol.* **2008**, *3*, 210.
[29] S. Pisana, M. Lazzeri, C. Casiraghi, K. S. Novoselov, A. K. Geim, A. C. Ferrari, F. Mauri, *Nat. Mater.* **2007**, *6*, 198.
[30] M. Lazzeri, F. Mauri, *Phys. Rev. Lett.* **2006**, *97*, 266407.
[31] L. M. Malard, M. A. Pimenta, G. Dresselhaus, M. S. Dresselhaus, *Phys. Rep.* **2009**, *473*, 51.
[32] A. C. Ferrari, J. C. Meyer, V. Scardaci, C. Casiraghi, M. Lazzeri, F. Mauri, S. Piscanec, D. Jiang, K. S. Novoselov, S. Roth, A. K. Geim, *Phys. Rev. Lett.* **2006**, *97*, 187401.
[33] Z. H. Ni, Y. Y. Wang, T. Yu, Y. M. You, Z. X. Shen, *Phys. Rev. B* **2008**, *77*, 235403.
[34] J. M. Cowley, J. A. Ibers, *Acta Crystallographica* **1956**, *9*, 421.
[35] R. Schlogl, W. Jones, H. P. Boehm, *Synth. Met.* **1983**, *7*, 133.
[36] T. Sasa, Takahash.Y, T. Mukaibo, *Carbon* **1971**, *9*, 407.
[37] L. Liu, Z. X. Shen, *Appl. Phys. Lett.* **2009**, *95*, 252104.
[38] H. Schmidt, T. Ludtke, P. Barthold, E. McCann, V. I. Fal'ko, R. J. Haug, *Appl. Phys. Lett.* **2008**, *93*, 172108.
[39] S. R. Su, D. W. Oblas, *Carbon* **1987**, *25*, 391.
[40] I. I. Mazin, A. V. Balatsky, *arXiv:0803.3765v1*.
[41] Z. H. Ni, H. M. Wang, J. Kasim, H. M. Fan, T. Yu, Y. H. Wu, Y. P. Feng, Z. X. Shen, *Nano Lett.* **2007**, *7*, 2758.

[42] Y. Y. Wang, Z. H. Ni, Z. X. Shen, H. M. Wang, Y. H. Wu, *Appl. Phys. Lett.* **2008**, *92*, 043121.
[43] J. P. Perdew, K. Burke, Y. Wang, *Phys. Rev. B* **1996**, *54*, 16533.
[44] G. Kresse, J. Furthmuller, *Comput. Mater. Sci.* **1996**, *6*, 15.

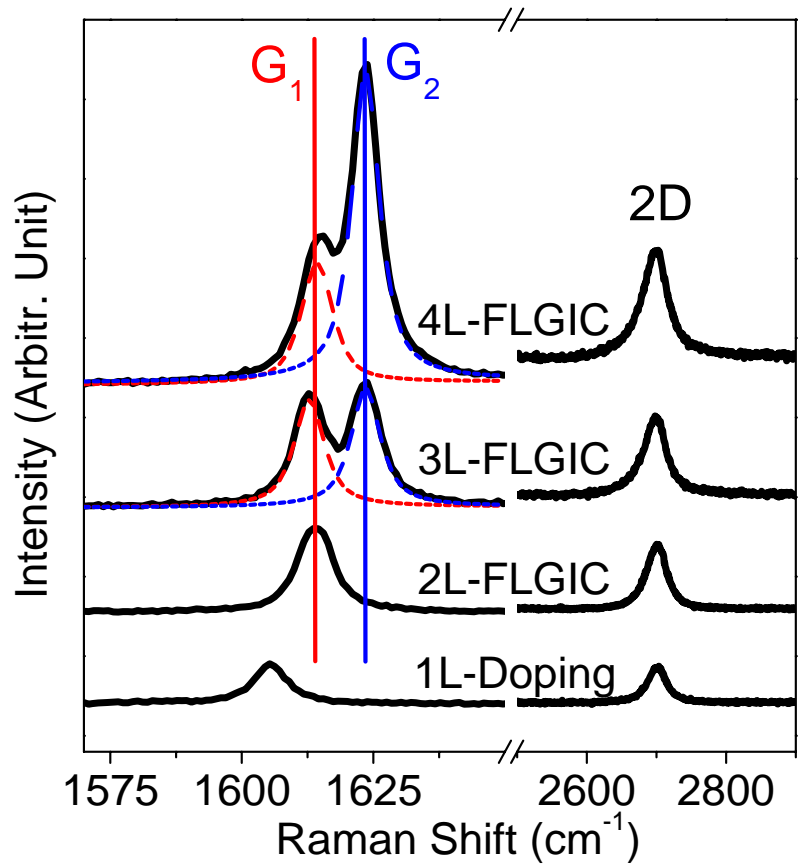

**Figure 1.** The Raman Spectrum of 1L doped graphene and 2L- to 4L- FLGIC. All the spectra were measured under same experimental conditions. The 2L-FLGIC shows a singlet G peak ($G_1$) at ~ 1612 cm$^{-1}$, while 3L- and 4L- FLGIC show doublet G peaks ($G_1$ and $G_2$ peaks locate at ~ 1612 cm$^{-1}$ and ~ 1623 cm$^{-1}$, respectively) with different intensity ratio. The $G_1$ and $G_2$ peaks are fitted with dashed red and blue curves, respectively. Each spectra shows single Lorentzian 2D peak and the intensity increases with the number of layers.

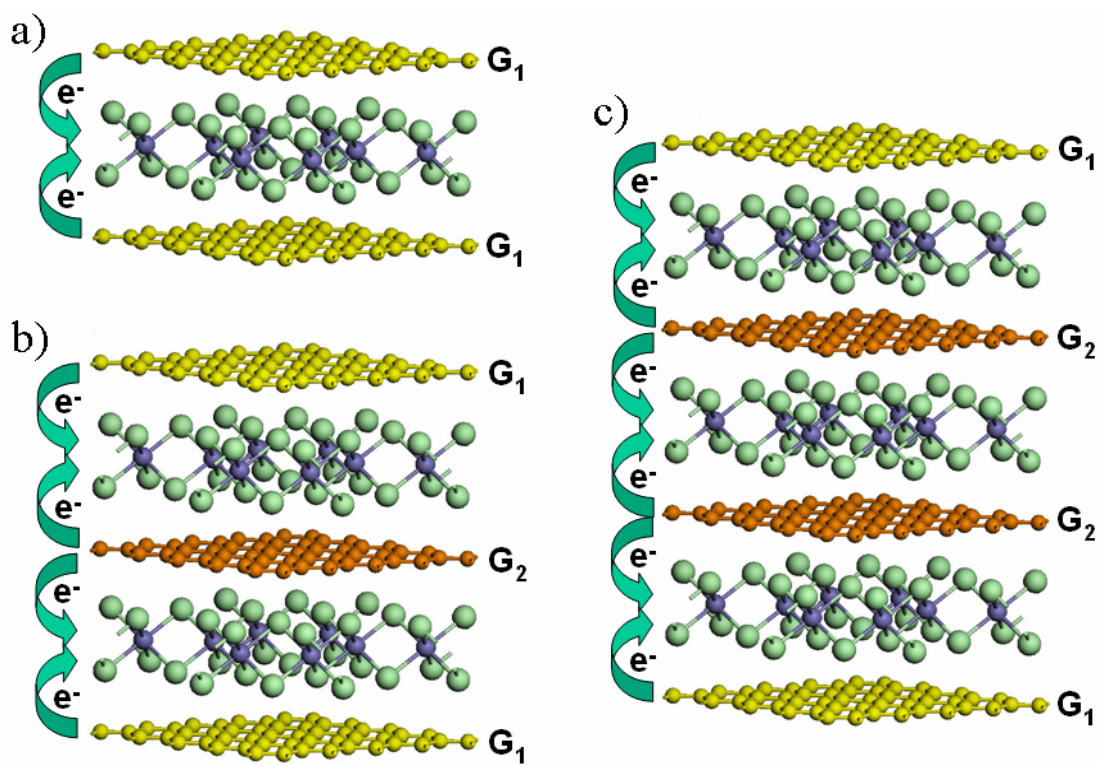

**Figure 2.** Schematic crystal structures of FLGIC. a) 2L-FLGIC. b) 3L-FLGIC. c) 4L-FLGIC. The model is constructed based on FeCl$_3$-GIC.[14, 22, 34] The graphene layers flanked on one/both side(s) by FeCl$_3$ layer(s) are denoted as yellow/orange. Cl atoms and Fe atoms are denoted as green and blue, respectively.

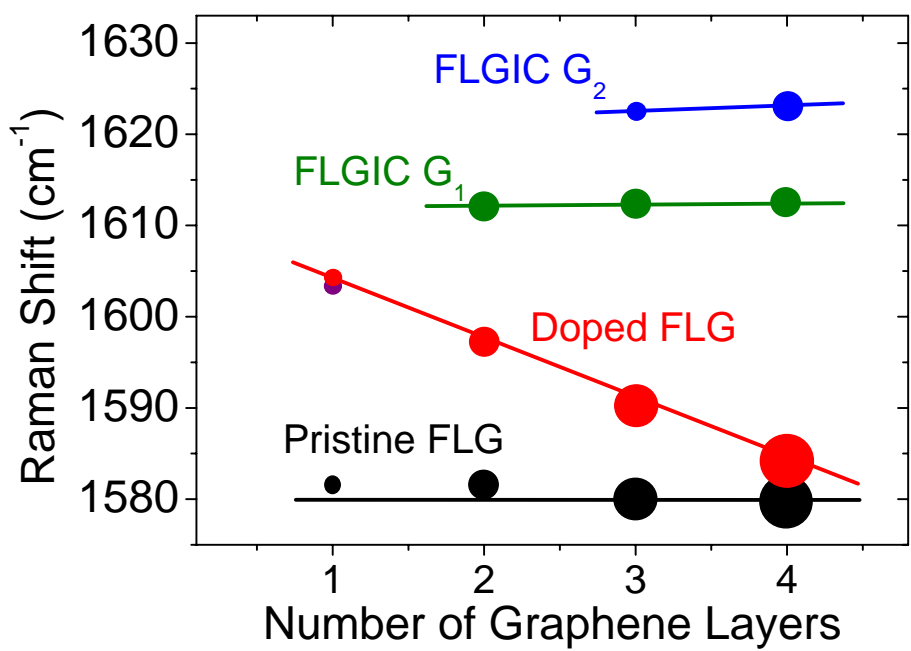

**Figure 3.** Comparison of the G peak position of FLGIC (green and blue), doped FLG (red) and pristine FLG (black). The SLG after intercalation process has similar spectrum as doped SLG (purple). The intensity of G peak is represented by different radius of the solid circles.

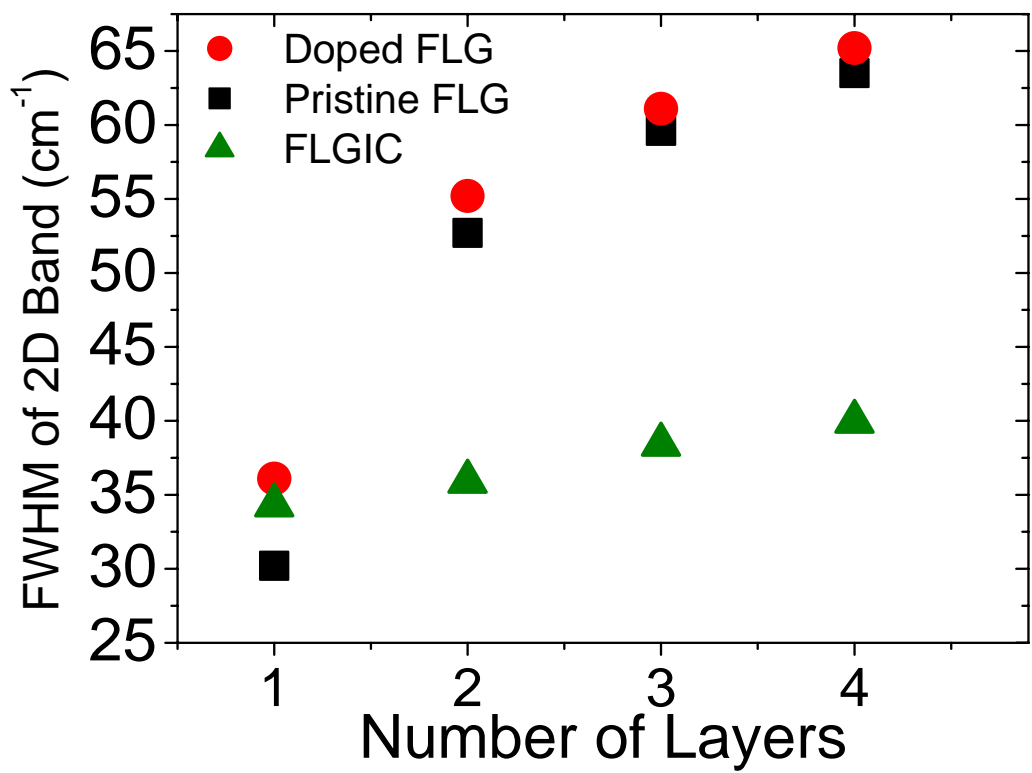

**Figure 4.** FWHM of the 2D peak of doped FLG (red circles), pristine FLG (black squares), and FLGIC (green triangles), where doping instead of intercalation for 1L case.

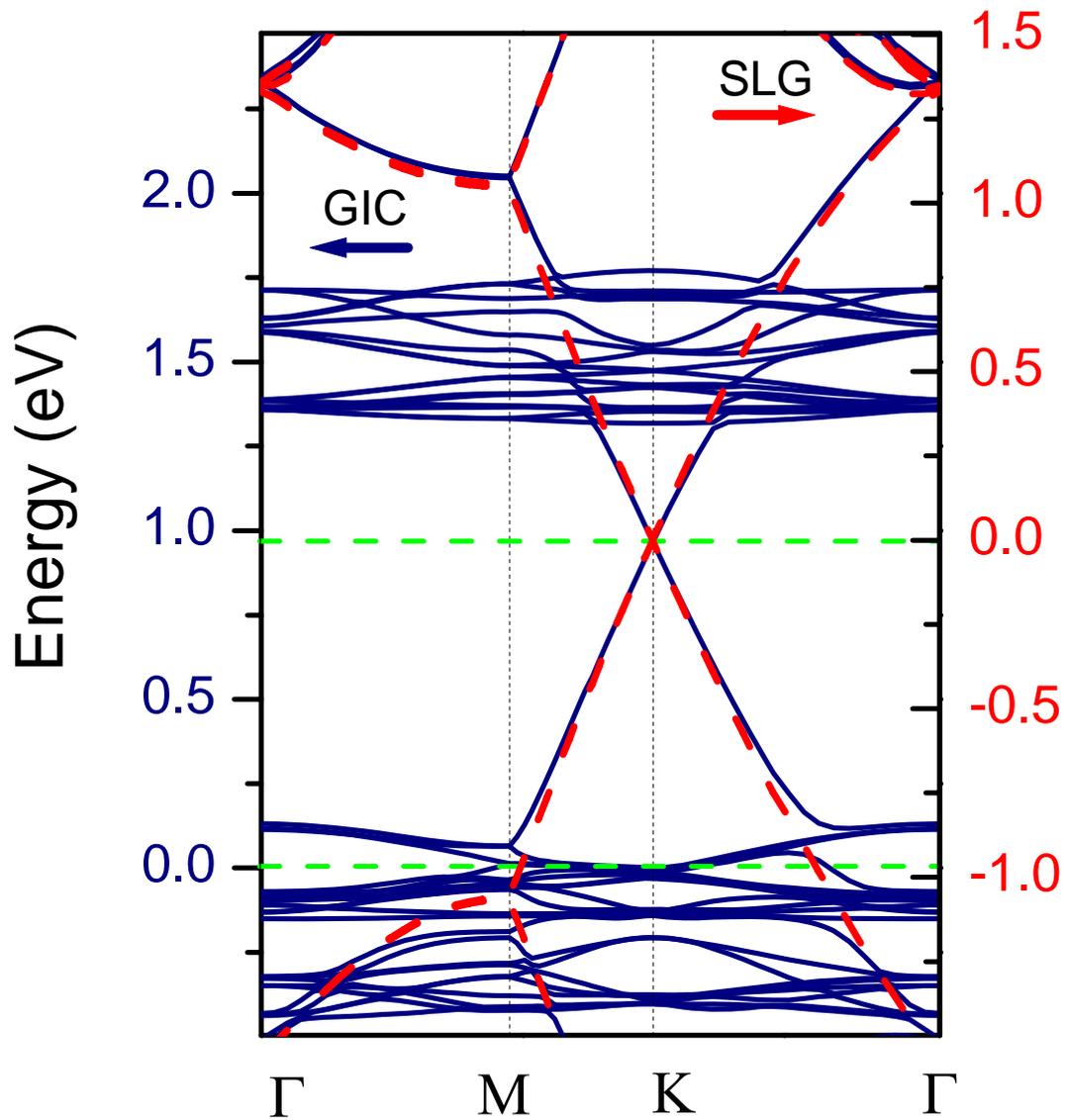

**Figure 5.** Electronic band structures of SLG (red dashed line) and FeCl$_3$-based stage 1 GIC (blue solid line). The horizontal bands of GIC originate from the *d* orbital of iron. Except for those bands, the band structure of GIC and SLG are very similar. An obvious difference is that the Dirac point of SLG locates exactly at Fermi surface while that of GIC locates at ~1 eV, suggest that GIC is strongly hole doped.

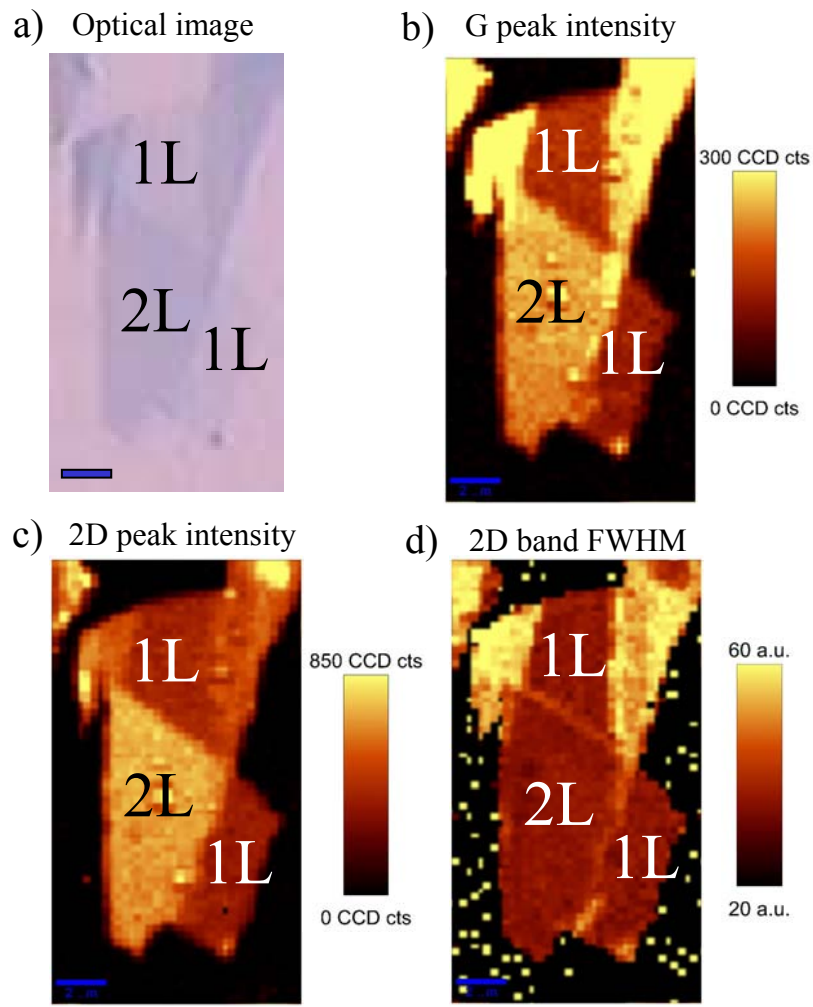

**Figure 6.** Raman images of 2L-FLGIC and 1L-graphene. a) Optical image of graphene samples before intercalation. b) Raman image of the G peak intensity after intercalation. c) Raman image of the 2D peak intensity after intercalation. d) Raman image of the 2D peak FWHM (linewidth) after intercalation. The intensities of G and 2D peaks of 2L-FLGIC are much stronger than those of 1L graphene, while no noticeable difference in the 2D linewidth between 2L-FLGIC and 1L-graphene. The 2D linewidth of 2L-FLGIC region is uniform at ~35 cm$^{-1}$. The scale bar is 2 μm for all images.

# Table of Contents Graphic

Iron chloride intercalated few layer graphene are successfully prepared and systematically studied by Raman spectroscopy. Raman spectra of such few-layer graphene intercalation compounds (FLGIC) clearly reveal the single-layer graphene like electronic structure and strong charge transfer induced doping effect. Such properties are further confirmed by first principle calculations. The successful fabrication of FLGIC opens a new way to modify properties of graphene for future applications.

Keyword: Graphene, Intercalation, Raman, Electronic band structure, Charge transfer, Doping

D. Zhan, L. Sun, Z. H. Ni*, L. Liu, X. F. Fan, Y. Y. Wang, T. Yu, Y. M. Lam, W. Huang, Z. X. Shen*

**$FeCl_3$ based Few-Layer Graphene Intercalation Compounds: Single Linear Dispersion Electronic Band Structure and Strong Charge Transfer Doping**

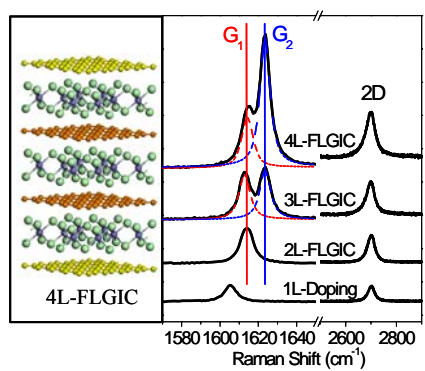